\documentclass[submission,copyright,creativecommons]{eptcs}
\providecommand{\event}{TARK 2023} 

\usepackage{iftex}

\ifpdf
  \usepackage{underscore}         
  \usepackage[T1]{fontenc}        
\else
  \usepackage{breakurl}           
\fi

\usepackage{comment}

\title{A ``Game of Like'' : Online Social Network Sharing As Strategic Interaction}
\author{Emmanuel J. Genot\thanks{I wish to thank Erik Mohlin, Jens Ulrik Hansen, Justine Jacot, and Patricia Rich, for their invaluable help at the various stages of this paper's development; three anonymous referees,  whose comments and suggestions brought about some major changes and (hopefully) improvements; and Rineke Verbrugge, who reviewed those changes, and suggested further improvements. Any mistakes left are on me.}
\institute{Department of Philosophy\\
Lund University\\
Lund, Sweden}
\email{emmanuel.genot@fil.lu.se}
}
\def\titlerunning{A Game of Like}
\def\authorrunning{E.J. Genot}

\newcommand{\like}{\textsf{\upshape like}}
\newcommand{\share}{\textsf{\upshape share}}

\newcommand{\se}{nline search engine}

\newcommand{\sn}{nline social network}
\newcommand{\osn}{\textsc{osn}}
\newcommand{\set}[1]{\{#1\}}

\newcommand{\eqref}[1]{(\ref{#1})}

\begin{document}
\maketitle

\begin{abstract} 
We argue that behavioral science models of online content-sharing overlook the role of strategic interactions between users. Borrowing from accuracy-nudges studies decision-theoretic models, we propose a basic game model and explore special cases with idealized parameter settings to identify refinements necessary to capture real-world o\sn \ behavior. Anticipating those refinements, we sketch a strategic analysis of content amplification and draw a connection between Keynes' ``beauty contest'' analogy and recent social-epistemological work on echo chambers. We conclude on the model's prospects from analytical and empirical perspectives.

\end{abstract}


\section{Motivations}\label{sec:Motivations}
O\se s garnered attention from social epistemologists in the early days of the commercial Internet, when A. Goldman analyzed them as retrieval systems in \cite{Goldman1999}. Later, T.A. Simpson extended Goldman's analysis into a model of surrogate expertise in \cite{Simpson2012} in direct response to Google Search personalization algorithms. Recently, epistemologists have turned to o\sn s (hereafter \osn), fulfilling a similar function of online information sources, with even greater personalization. Notably, T.C. Nguyen \cite{Nguyen2020} provided a much-needed conceptual analysis of \osn\ epistemic bubbles and echo chambers, and C. O'Connor and J.O. Weatherall \cite{OConnor2019} proposed that applying network epistemology to \osn\ could address limitations of contagion models of online information spread. At the same time, behavioral scientists have independently addressed the limitations of contagion models by looking at \osn-sharing through a rational choice lens. Particularly, studies that shaped the field and its public perception have manifested a Bayesian influence. Widely publicized studies like \cite{Vosoughi2018} (a \textit{Science} cover story: ``How lies spread--On social media, fake news beats the truth") and \cite{Pennycook2021} (a \textit{Nature} cover story: ``Misinformation--A prompt to check accuracy cuts online sharing of fake news") appealed to Bayesian decision theory and expected utility theory to rationalize \osn\ content-sharing and interpret diffusion-model data analyses.\footnote{%
``[U]ser characteristics and network structure could not explain the differential diffusion of truth and falsity, we sought alternative explanations for the differences in their diffusion dynamics. One alternative explanation emerges from information theory and Bayesian decision theory. Novelty attracts human attention, contributes to productive decision-making, and
encourages information sharing because
novelty updates our understanding of the world.'' \cite[p. 1149]{Vosoughi2018}.  Similarly,  ``people do care more about accuracy than other content dimensions but accuracy nonetheless often has little effect on sharing, because (ii) the social media context focuses 
[users'] attention on other factors such as the desire to attract and please followers/friends or to signal one’s group membership. In the language of utility theory, an `attentional spotlight' is shone upon certain terms in the decider’s utility function, such that  only those terms are weighed when making a decision''  \cite[p. 591]{Pennycook2021}. The framework of \cite{Pennycook2021} is implicitly decision-theoretic, as utilities take as argument proxies for individual choices (content shared) rather than strategic profiles (cf. Section \ref{sec:GameOfLike}).\label{no:NoveltyAttention}} %

Decision theory best models decisions {under uncertainty about the state of nature}, but \osn-sharing outcomes depend on reactions from a community of users. The preferred model for decisions \textit{under uncertainty about other agents' decisions} is game theory, and while the formalisms are inter-translatable, decision theory is less expressive. As pointed out by J. Harsanyi, the game-to-decision direction loses in translation the explicit expression of mutual expectations of rationality (via solution concepts \cite{Harsanyi1982a,Harsanyi1982b}). Compounding this issue, decision-theoretic models from behaviral science studies (such as \cite{Vosoughi2018,Pennycook2021}) were not proposed as translations for games and thus did not explicitly translate mutual expectations into constraints on decision-makers priors (as per the games-to-decision direction, cf. \cite{KadaneLarkey1982a,Harsanyi1982a}), leaving their  role almost entirely unanalyzed. Unfortunately, social epistemology offers no ready-made solution. Nguyen's analysis is strategic but informal and cannot bear on the data without a formal reconstruction, while network epistemology does not address strategic expectations formally.

The absence of a strategic analysis of \osn-sharing motivated the approach presented in the remainder of this paper. Section \ref{sec:GameOfLike} builds upon behavioral science decision-theoretic models to propose a simplified game model for \osn-sharing, differentiating between content-based and engagement-based preferences. Section \ref{sec:StrategySelection} examines special cases that highlight the model's salient features and limitations and identifies extensions necessary to reconstruct real-life \osn\ users' behavior. Section \ref{sec:MutualExpectation} extrapolates informally and proposes that special cases of \osn-sharing elicit strategy selection akin to reasoning in guessing games and could illuminate content amplification scenarios, including Nguyen's epistemic bubbles and echo chambers. We conclude with the analytic prospects of a strategic re-interpretation of extant data, and a suggestion for the design of new studies.

\section{A Game of Like}\label{sec:GameOfLike}

Behavioral science studies of \osn-sharing often acknowledge the role of strategic interactions between users but have so far fallen short of factoring in their contribution. Pennycook \textit{et al.} (2021) is a paradigmatic example: the authors note that ``the desire to attract and please followers/friends or to signal one’s group membership''  \cite[p. 591]{Pennycook2021} contributes to content-sharing decisions, but propose a utility function limited to personal preferences for content having such-and-such characteristics.\footnote{%
``Consider a piece of content $x$ which is defined by $k$ different characteristic dimensions 
[including] whether the content is false/misleading $F(x)$, and other $k-1$ dimensions [that] are non-accuracy-related (e.g. partisan alignment, humorousness, etc) defined as $C_2(x)\dots C_k(x)$. In our model, the utility 
[\dots] from sharing content $x$ is given by: $U(x)=-\alpha_1\beta_F F(x)+ \sum^k_{i=2}\alpha_i\beta_i C_i(x)$ where $\beta_F$ indicates how much they dislike sharing misleading content and $\beta_2\dots\beta_k$ indicate how 
much they care about each of the other dimensions (i.e. $\beta$s indicate preferences); while $\alpha_1$
indicates how much the person is paying attention to accuracy, and $\alpha_2\dots\alpha_k$ indicate how much the person is paying attention to each of the other dimensions.'' \cite[Supplementary Information: S9-10]{Pennycook2021} \label{no:PersonalUtility}} %
A natural first step toward a strategic model is thus to introduce the missing terms, then specify a game based on this completed picture. For simplicity, we can let $u_{p_i}(\cdot)$ denote $i$'s \textit{personal utility}, expressing how some content aligns with $i$'s personal preferences for content having such-and-such characteristics, with the understanding that this alignment could be further analyzed along multiple dimensions (as in \cite{Pennycook2021}, cf. n. \ref{no:PersonalUtility}).
To that, we add a term that we denote  $u_{s_i}(\cdot)$, for the \textit{social utility}, expressing how reactions to the content shared---`likes,' re-shares, comments, etc.---satisfy $i$'s preferences for social validation or, more generally, engagement.  Finally, we introduce a parameter, that we denote $\gamma$, to represent the relative weights of $i$'s personal preferences for content and social preferences for engagement. 
In the decision-theoretic model of \cite{Pennycook2021}, the only action being `sharing,' actions and content are indiscernible, and the utility function can range over the content. In a game-theoretic model, the utility function ranges over \textit{strategy profiles}, and we must distinguish content from actions. 

As a basic model, we consider an $n$-player repeated game $G$ in strategic form with a set $P$ of players, where any $i\in P$ can, at each round, `like' or `share' content. As simplifications, we assume that players only share {new} content at round $r=0$, so any `share' action at round $r\geq 1$ is a `reshare.' Under this simplification, we can specify a set of (original) \textit{content} $C_G=\set{c_1,\dots,c_n}$  for $G$, where $c_i$ is the content introduced at $r=0$ by agent $i$. The \textit{action set} for some player $i$ at some round $r$ is $A_i=\set{\like_i(x,y),\share_i(x,y): x\in C_G, y\in P}$, where $y$ is a player who shared $x$ at some round $r'<r$, and from whom $i$ is re-sharing $x$. Note that, under our simplifying assumption, at $r=0$, there is nothing to `like.' If all actions are visible to all players, no restriction is imposed on $x$ or $y$. Explicitly: any content shared by some player at round $r$ can be reshared by any other player at round $(r+1)$. This amounts to a game with perfect information, adequate for demonstrating the strategic standpoint's fruitfulness but insufficient to model real-world \osn s (see below). Our earlier discussion of personal and social preferences yields a utility function, as below.
\begin{eqnarray}
    U_{i}(\cdot) & = & \gamma_i u_{p_i}(\cdot) + (1-\gamma_i)u_{s_i}(\cdot)
    \label{eq:Utilities}
\end{eqnarray}
Intuitively, in the decision-theoretic approach, $u_{p_i}(c)$ expresses $i$'s preferences as a function of the distance between $c$ and $i$'s ideal content located in a multi-dimensional space whose dimensions correspond to $i$'s criteria of evaluation. In a round of $G$,  the argument of \eqref{eq:Utilities} is a strategy profile $\overline\sigma=(\sigma_1,\dots \sigma_n)$ where $\sigma_i$ is player $i$'s strategy at that round.  Following the same intuition, $u_{p_i}(\overline\sigma)$ can be thought of as a function of the relative distances between $i$'s ideal content and the  `community content' $c_1, \dots, c_n$, or some weighed sum thereof, representing how close $C_G$ is to $i$'s ideal.\footnote{%
Note that $i$ may be indifferent to others' strategies, in which case $u_{p_i}(\overline\sigma)=u_{p_i}(\overline\sigma')$ whenever $i$'s strategy $\sigma_i$ is the same in $\overline\sigma$ and $\overline\sigma'$.} %
So construed, and under our simplification, $u_{p_i}(\overline\sigma)$  remains constant after $r=0$. Again, this personal preference model is sufficient for our purposes. Still, in a real-world \osn, overall engagement could indirectly affect $u_{p_i}(\overline\sigma)$ ($i$ may care for overall visibility, and in a model with incomplete information, visibility would depend on engagement, see below).

In a decision-theoretic model (e.g., extending that of \cite{Pennycook2021}) $u_{s_i}(c)$ would be a function of the (accumulated) engagement from users other than $i$, with $c$ (when shared by $i$). In $G$, $u_{s_i}(\overline\sigma)$ at round $r$ is, in part, a function of how other players have engaged in $r$ with the content $i$ shared at some $r'>r$; and in part, of the accrued social utility inherited from earlier rounds. The candidate functions for computing either component are too numerous to review here, and which one applies to particular cases may be empirically constrained by algorithms. 
Still, it suffices for our purposes to note that, at some round $r$,  $u_{s_i}(\overline\sigma)$ does not `reset' $i$'s social utility; that the contribution of `likes' and (re)shares may vary; and that evaluations may depend on players' knowledge.\footnote{%
For a concrete example, Twitter's ranking algorithm weighs `like' reactions more than re-tweets (reshares) when determining which content should appear in users' feeds. A knowledgeable user may prioritize sharing content they believe would receive `likes' to optimize the chances that other users are exposed to their content later, whether they value engagement as social validation or as a means to increase content visibility.} %
For definiteness, we can assume a function $u_{s_i}(\overline\sigma)$ that ranks higher strategy profiles where content $i$ shared (or reshared) is both {liked and reshared} rather than liked or shared (alone)---i.e., a function that takes some weighed sum of `likes' and (re)shares, rather than an average (or an argmax).  This justifies the shorthand ``{game of like}"---as a nod to J. Conway's ``game of life'' \cite{Gardner1970}---since the preferred social outcome, over repetitions, is like-and-reshare, a strengthened form of `like' (``game of share'' would be equally justified, but the homage and homophony would be lost).

Let us conclude this section with a few words on our model's (self-imposed) limitations. In real-world \osn,  new content can be introduced at any time, and  players have only a partial picture of the content they can reshare. A more realistic ``game of like'' would have {imperfect information} (e.g., as a model of bounded attention): any content $c$ would be available to a player $i$ to react to at $r$ with a certain probability, depending  on overall engagement with $c$ prior to $r$. In such a model, $i$ could be aware of some $c$, close to $i$'s ideal content, and care for its visibility (the probability of $c$ being available to other players) and thus for other players' engagement with $c$. Conversely, $i$ might not worry much about some $c$, far removed from their ideal, as long as $c$'s probability of being available to other players would remain low. Still, a simplified model with perfect information already acknowledges the relevance of overall interaction by virtue of the argument of $\gamma_i u_{p_i}(\cdot)$ being a strategy profile, and thus furthers goal of identifying strategic components of \osn-sharing. Hence, our ``game of like'' with perfect information is a proof-of-concept and a foundation for future developments. The next section considers special cases, varying players' $\gamma$ types, to determine which refinements would be necessary to turn the proof-of-concept model into a model for real-world \osn s.


\section{Strategy Selection}\label{sec:StrategySelection}

Let us begin with the limit case where, for all $i\in P$, $\gamma_i=1$, denoted $G_{\gamma=1}$ for later reference. We could distinguish \textit{a priori} between a variety of subcases, depending on whether players have non-equivocal prior beliefs about other players' personal preferences; and/or whether they have non-equivocal prior beliefs about other players' $\gamma$. However, the differences between those subcases are inconsequential. To see this, assume an arbitrary player $i$ in $G_{\gamma=1}$ who \textit{does have} non-equivocal prior beliefs about other players' personal preferences for content and $\gamma$-type (say, following a round of cheap talk). \textit{Ex hypothesis}, at any round $r$ of $G_{\gamma=1}$, for any $i\in P$, $U_i(\overline\sigma)$=$u_{p_i}(\overline\sigma)$. Hence, $i$'s best strategy at round $r=0$ is to share whatever content $c_i$ available to them that is closest to their ideal content (according to their dimensions of evaluation). Beliefs about other players' preferences and $\gamma$ type do not affect that choice. Hence, $i$ would choose the same content \textit{without} any information about other players. Since the only assumption we made about $i$ is that $\gamma_i=1$, this generalizes to any $i\in P$ for $G_{\gamma=1}$ (and yields an  equilibrium solution in the basic model for $r=0$ in $G_{\gamma=1}$).
Under the assumption that content is only introduced at round $r=0$, the distance between the `community content' and any player $i$'s ideal content remains constant across repetitions, whatever their strategy at $(r\geq 1)$.  Relaxing this simplifying assumption is one way to model how players can attempt to drive community content closer to their preferences by sharing more content closer to their ideal at any new round $(r\geq 1)$. But this would not bring the model closer to real-world \osn, as ``spamming''  content is only efficient if the content is visible, bringing us back to a version of the ``game of like'' with imperfect information. Conversely, a ``game of like'' \textit{without} content introduced at round $r>0$, and with $(\gamma=1)$-players only, would be susceptible to manipulations by coalitions of like-minded players, who would want to see some content promoted. Therefore, relaxing the assumption that no new content is introduced past $r=0$ would not be especially illuminating without an explicit topological model of content distances and auxiliary assumptions about how variable availability of content correlates with engagement.

In a second limit case,  denoted $G_{\gamma=0}$, all $i\in P$ are such that $\gamma_i=0$. Unlike $G_{\gamma=1}$, player priors about others can significantly impact the game. To see this, consider the limit subcase where players' $\gamma$ type is common knowledge. Then, $G_{\gamma=0}$ becomes a game of reciprocation-or-retaliation or \textit{quid pro quo}, where players either trade reciprocal `likes' and re-shares, or ignore one another, and where content becomes inconsequential (so that it matters little whether new content can be introduced after $r=0$ or not). 
To see this, consider a simple $G_{\gamma=0}$ case with two players $i$ and $j$, content introduction restricted to $r=0$, and (as a simplification) no marginal utility for `liking' or re-sharing one's content. Hence, the only utility $i$ and $j$ can get is from the other player's liking or re-sharing their content. 
At $r=0$, they share (resp.) $c_i$ and $c_j$. At $r=1$, $i$ ($j$) can like or re-share $c_j$ ($c_i$), or 
do nothing (for definiteness: repeating their move from $r=0$). If either does nothing at $r=1$, the other can retaliate at $r=2$ by playing nothing; otherwise, they can reciprocate and play the remaining action (like, or reshare) they did not play at $r=1$. With no introduction of new content, they can repeat the cycle over $c_i$ and $c_j$. If new content is allowed, they can repeat cycles of three rounds (introduction, like, or re-share, then reciprocation or retaliation) to accrue utility. The strategy just described turning the ``game of like" into a game of reciprocation-or-retaliation, and resembles the \textit{tit-for-tat} strategy in the repeated Prisonner's Dilemma. 

As extreme as it is, this case suggests that when $(\gamma=0)$-players have non-equivocal beliefs about one another's $\gamma$ type, the closer the players are to having correct beliefs, the closer $G_{\gamma=0}$ resembles a \textit{quid pro quo} game. Assume now a subcase of $G_{\gamma=0}$ where players have equivocal beliefs about $\gamma$ types---i.e., do not \textit{know} that other players are $(\gamma=0)$-players. If they also have equivocal beliefs about other players' personal preferences for content, the rational choice (for any $i$) is a mixed strategy assigning equal weight to any content $i$ has access to at $r=0$ and hope for the best. Lifting the restriction on content introduction is more consequential than in the $G_{\gamma=1}$ case, as repeated observations of others' sharing behavior are necessary to infer their personal preferences for content from their actions or their preferences for engagement. Since, \textit{ex hypothesis}, no player in $G_{\gamma=0}$ actually cares for content (as long as they receive engagement), inferences from sharing behavior to personal preferences could result in `false consensus' situations if players gradually amplify a salient type of content, leading to an echo chamber (in the sense of \cite{Nguyen2020}; cf. Section \ref{sec:MutualExpectation}). However, even without lifting the assumption, we can form a picture of a repeated game with new content by assuming a round of cheap talk prior to $r=0$, during which players can form priors (or update equivocal priors) about other players' preferences based on observed behavior. Suppose that some candidate content appears salient for eliciting positive reactions---say,  pictures  of cats in precarious positions. Then, upon engaging in  $G_{\gamma=0}$, players could anticipate similar pictures to elicit `like' and `share' reactions, skewing the content shared at $r=0$ toward pictures of cats in precarious positions. Thus, it would appear that a majority of players favor cat pictures. Even without the introduction of new content, this could lead to cat pictures being increasingly reshared at every $r\geq 1$  without (\textit{ex hypothesis}) any player selecting their strategy out of personal preference for that type of content, resulting in a `false consensus.' Again, as with $G_{\gamma=1}$, how engagement could impact visibility appears more critical than whether or not content may be repeatedly introduced. 
Subsequently, the need to accommodate $(\gamma<1)$-players does not require further refinements beyond those suggested by $G_{\gamma=1}$: imperfect information and an explicit  content evaluation and comparison model. The latter would, in particular, suffice for representing how $(\gamma<1)$-players form (and revise) beliefs about the majority's opinion, instrumental in selecting strategies for eliciting engagement.

\section{Mutual Expectations and Social Influence}\label{sec:MutualExpectation}

Our remark about the majority's opinion being of import to $(\gamma<1)$-players may remind the reader of J.M. Keynes' ``Beauty Contest'' analogy for professional investment, quoted below.
\begin{quote}
[P]rofessional investment may be likened to those newspaper competitions in which the competitors have to pick out the six prettiest faces from a hundred photographs, the prize being awarded to the competitor whose choice most nearly corresponds to the average preferences of the competitors as a whole. [\dots] It is not a case of choosing those [faces] that, to the best of one's judgment, are really the prettiest, nor even those that average opinion genuinely thinks the prettiest. We have reached the third degree where we devote our intelligences to 
anticipating what average opinion expects the average opinion to be. 
\cite[p. 156]{Keynes1978} 
\end{quote}
The parallel is intentional: we propose that Keynes' ``third degree'' describes the reasoning of a $(\gamma<1)$-player selecting a strategy that could elicit (re)share reactions from other $(\gamma<1)$-players who would want to elicit `like' reactions. More generally, a ``game of like'' with some proportion of $(\gamma<1)$-players relates to \textit{guessing games}, proposed as a generalization of J.M. Keynes' beauty contest by R. Nagel (first, in \cite{Nagel1995}; see \cite{NagelBosh2002} for an overview of empirical studies). 
A formal reconstruction of this suggestion would require an explicit model of preference distances (already identified as a necessary refinement for our basic model to capture real-world \osn s), but we can offer an informal sketch.

Asssume the standpoint of a player of type $\gamma=0$, that we will denote $\gamma_0$, reasoning about other players of a ``game of like."\footnote{%
We assume that the agent is a $(\gamma=0)$-player rather than a weaker $(\gamma<1)$ to avoid dealing with correlations between personal preferences for content and preferences for engagement. Otherwise, we would have to factor in the cost of sharing contrary-to-preference content, which could offset the benefit of engagement.} %
When choosing between multiple options for content to share, when $\gamma_0$'s goal is accruing ``like''  reactions, $\gamma_0$ is equally well-off: ($i$) choosing based on their own preferences for content, or: ($ii$) choosing based on the majority's preference (e.g., as inferred following a round of cheap talk) when preferences agree; and: ($iii$) possibly worse off, when preferences disagree. In case ($iii$), $\gamma_0$ would be better off switching to an option that agrees with the majority's (displayed) preferences. Thus, options based on $\gamma_0$ preferences are \textit{weakly dominated} by options based on the majority's preferences (as inferred by $\gamma_0$).
Consider now how $\gamma_0$ would approach selecting a strategy for eliciting ``share'' reactions; as simplification, assume that $\gamma_0$ believes that most players are like him and care more for engagement than for content. Then, $\gamma_0$ expects that most players would (re-)share content to elicit (at least) `like' reactions. If $\gamma_0$ assumes that those players are rational, they expect those players to reason to ($i$--$iii$) above. From there, $\gamma_0$ can conclude that selecting an option based on their own preferences for content would yield the same payoff as choosing based on the majority's opinion of the majority's (displayed) preferences for content (if in agreement); and possibly a worse payoff (if in disagreement). In the latter case, $\gamma_0$ would be better off switching options. Hence, a selection based on the majority's opinion of the majority's (displayed) preferences \textit{weakly dominates} a selection based on $\gamma_0$'s preferences for content alone.

The argument just sketched guesstimates too many important parameters to be general---e.g., the respective distribution of $\gamma$ types among the players, the cost of seeking social feedback with contrary-to-personal preferences for other players that $\gamma_0$, how $\gamma_0$ would arrive at estimates for those, etc. However, it suffices to motivate a comparison between a subclass of ``game of like,''  Keynes' beauty contest, and Nagel's guessing games. And empirical motivation for this comparison would be the reconstruction of the real-world \osn\ behavior colloquially called `signal boosting,' whereby users of an \osn\ leverage the influence of public figures (``influencers") with a larger following base, tagging them in hope to be re-shared. A well-reported example is a November 13, 2020 Twitter video featuring actor R. Quaid reading aloud an earlier tweet from then-US president D.J. Trump under a stroboscopic light, with an over-dramatic voice. Trump  (unsurprisingly) reshared Quaid's video, which then accrued millions of views from Trump's followers, reaching beyond Quaid's following. In fact, we have already encountered in Section \ref{sec:StrategySelection} a variant of (involuntary) signal-boosting behavior, as a pathway to amplification (false consensus) when discussing $G_{\gamma=0}$. This seems grounds enough to suggest that a ``game of like'' model of \osn\ with influencers could contribute to a formal theory of online amplification, echoing Keynes’ motivations for the beauty-contest analogy (speculative asset bubbles). 

Another possible contribution that circles back to social epistemology is a possible formal reconstruction of Nguyen's conceptual analysis \cite{Nguyen2020}. Nguyen proposes that \textit{epistemic bubbles} occur when individuals receive limited exposure to information sources challenging their pre-existing beliefs, in contrast to \textit{echo chambers}, which emerge when individuals receive extensive exposure to information sources that align with their pre-existing beliefs. Epistemic bubbles result from combined personal choice and algorithmic curation, particularly when online platforms tailor content to individual preferences, thereby restricting the information individuals encounter. In an echo chamber, people reinforce their views and are shielded from diverse perspectives and alternative information, leading to the exclusion of dissenting opinions. Nguyen notes that epistemic bubbles are easy to burst with the presentation of contrary evidence, while echo chambers are self-reinforcing, with social interaction actively fostering distrust of outside sources. Nguyen's analysis of echo chambers invites a formal reconstruction in a ``game of like'' model with imperfect information, bringing it closer to the methodological frameworks of behavioral science (\textit{modulo} a game-to-decision translation).

\section{Concluding Remarks}\label{sec:Conclusion}
We argued in Section \ref{sec:Motivations} that, while \osn-sharing is a strategic interaction, behavioral science models overlook the contribution of strategic anticipations. We extrapolated from behavioral science decision-theoretic models a basic game model of \osn-sharing (Section \ref{sec:GameOfLike}) and explored some limit cases to determine refinements necessary to capture real-world \osn-sharing (Section \ref{sec:StrategySelection}). A connection with Keynes' Beauty contest (and, more generally, guessing games) allowed us to sketch a strategic analysis of content amplification in the presence of influencers and users leveraging influence and suggested a direction for the model's development (Section \ref{sec:MutualExpectation}). Still, a ``game of like'' model may not contribute to conceptual analysis beyond a formal reconstruction of Nguyen's framework. And Nguyen's informal analysis has already done the heavy lifting of rigorously ordering concepts inherited from unsystematic public discourse, such as ``echo chambers'' and ``filter bubbles'' (introduced, resp., in \cite{Sunstein2001} and \cite{Pariser2011}), whose previously heterogeneous use had prevented consensus among researchers (see \cite{Ross2022}).
Rather, the litmus test for a ``game of like'' model would be a contribution to the critical re-evaluation of empirical data assessed from a decision-theoretic standpoint; and a suggestion of empirical investigations that a decision-theoretic standpoint would have neglected. To conclude, we want to suggest that, as incomplete as it is, our ``game of like'' model already achieves that.

As for critical re-evaluation, consider the widely-publicized study by Pennycook \textit{et al.} \cite{Pennycook2021}, in which the intervention condition proceeds from the auxiliary hypothesis that accuracy competes for attention with social incentives.\footnote{%
``In the control condition of each experiment, participants were shown 24 news headlines (balanced on veracity and partisanship) and asked how likely they would be to share each headline on Facebook. In the treatment condition, participants were asked to rate the accuracy of a single non-partisan news headline at the outset of the study (ostensibly as part of a pretest for stimuli for another study). They then went on to complete the same sharing intentions task as in the control condition, but with the concept of accuracy more likely to be salient in their minds.'' \cite[p. 591]{Pennycook2021}}
From a strategic standpoint, the authors' other auxiliary hypothesis---that ``people do care more about accuracy than other content dimensions'' (p. 591)---could characterize common knowledge of one dimension of users' preferences. If it does, having ``the concept of accuracy more [\dots] salient in [one's] mind'' (\textit{ibid}) could \textit{prime} engagement-based expectations, rather than shutting them down; in a game-to-decision translation, a Bayesian decision-maker would then anticipate a better prospect of eliciting other users' reactions conditional on being perceived as accurate (compared to conditional on being perceived as inaccurate). Compare this with the intervention condition from the more recent study by Ren \textit{et al.} \cite{RenDimantSchweitzer2023}, which socially incentivized both accuracy and engagement.\footnote{%
``In each incentive condition, we told participants that they would be entered into a lottery for a \$50 prize, and we manipulated how they would earn tickets to increase their odds of winning the prize.
In the \textit{Accuracy} condition, we told participants that they would earn a ticket if the post they shared was validated to be true by a professional fact-checker. In the \textit{Like} condition, we told participants that they would earn a ticket for each “like” they received from others. In the \textit{Comment} condition, we told participants that they would earn a ticket for each comment they received from others. In the \textit{Control} condition, we did not provide additional incentives. 
\cite[104421:4]{RenDimantSchweitzer2023}} %
%
As for the design of new studies, consider the question of whether differences in intervention conditions between \cite{Pennycook2021} and \cite{RenDimantSchweitzer2023} translate into differences in reasoning about other users' strategies is an interesting question. A positive answer would partition ``accuracy nudges'' into two classes (engagement-based and non-engagement-based). A negative answer would invalidate the auxiliary hypothesis that accuracy {competes} with the social dimension. The connection we drew with Nagel's work on guessing games suggest an empirical approach to answering this question, with following the methodology of \cite{CoricelliNagel2009}, which established neural correlates of lower- and higher-order ``Keynes degree'' reasoning in guessing games.

\nocite{*}
\bibliographystyle{eptcs}
\bibliography{biblio}
\end{document}